\documentclass[intlimits,twoside,a4paper]{article}

\usepackage{amsmath,amssymb}
\usepackage{graphicx}
\usepackage{wasysym}

\usepackage[T2A]{fontenc}
\usepackage[cp1251]{inputenc}

\usepackage{graphics}
\newcommand{\mean}[1]{\left\langle{#1}\right\rangle}
\newcommand{\Dt}{\frac{\mbox{d}}{\mbox{d}t}}

\newcommand{\dt}{\frac{\partial}{\partial t}}

\usepackage[eqsecnum]{cmpj2}

\issue{2016}{19}{3}{34001}
\doinumber{10.5488/CMP.18.34001}

\articletype{Rapid communication}

\title[Thermodynamics of voluntary behavior]%
{Linear non-equilibrium thermodynamics of human voluntary behavior:
a canonical-dissipative Fokker-Planck equation approach involving
potentials beyond the harmonic oscillator case}
\author[J.M. Gordon, S. Kim, T.D. Frank]{J.M. Gordon\refaddr{label1},
S. Kim\refaddr{label1},
T.D. Frank\refaddr{label1,label2}}
\addresses{
\addr{label1}
CESPA, Department of Psychology,
University of
Connecticut, 406 Babbidge Road, Storrs, CT 06269, USA\\
\addr{label2}
Department of Physics,
University of
Connecticut, 2152 Hillside Road, Storrs, CT 06269, USA
}
\date{Received June 7, 2016}
\authorcopyright{J.M. Gordon, S. Kim, T.D. Frank, 2016}

\begin{document}

\maketitle

\begin{abstract}
A novel experimental paradigm and a novel modelling approach are presented to
investigate oscillatory human motor performance by means of a key concept from
condensed matter physics, namely, thermodynamic state variables. To this end,
in the novel experimental paradigm participants performed pendulum swinging
movements at self-selected oscillation frequencies
in contrast to earlier studies in
which pacing signals were used. Moreover, in the novel modelling approach, a
canonical-dissipative limit cycle oscillator model was used with a
conservative part that accounts for nonharmonic oscillator components in
contrast to earlier studies in which only harmonic components were
considered. Consistent with the Landau theory of magnetic phase transitions, we
found that the oscillator model free energy decayed when
participants performed oscillations further and further away from the Hopf
bifurcation point of the canonical-dissipative limit cycle oscillator.

\pacs
05.40.Jc, %Brownian motion
 05.45.-a, %nonlinear dynamics and chaos
05.70.Ce, %thermodynamic functions and equations of state
05.70.Ln, %nonequilibrium and irreversible thermodynamics
64.70.qd, %thermodynamics and statistical mechanics
87.19.rs %Movement
\end{abstract}

In view of the success of physics in general and condensed matter physics
in particular, the question has been debated whether concepts from
physics and condensed matter physics can be generalized to describe not
only equilibrium systems but also non-equilibrium systems
(e.g.~\cite{glansdorff71book,haken77book}).
The canonical-dissipative (CD) approach to
 non-equilibrium
systems~\cite{haken73zeitphys,graham73a,ebeling04book,mongkolsakulvong10cmp}
in this context
plays an important role because it exhibits key  features of condensed matter
physics such as Hamiltonian mechanics and thermodynamic state
variables~\cite{frank05book}.
Moreover, in a series of recent experimental studies it has been
shown that the CD approach can be applied to characterize
the performance of humans --- as paradigmatic non-equilibrium
systems --- in simple oscillatory motor control
tasks~\cite{dotov11mc,dotov15biosystems,kimseokhun15}.
However, these earlier studies were subjected to two severe
limitations.
First, the experimental studies involved a pacing signal which is
inconsistent with the model of an autonomous limit-cycle oscillator.
Although in the study by Kim et al.~\cite{kimseokhun15},
the impact of the pacing signal was
greatly reduced, the question arises whether or not the
experimental design can be
improved to do without any pacing signal. Second, all three studies used
the standard CD oscillator model exhibiting the
conservative part of a harmonic oscillator. In this regard, the
question arises how to go beyond the harmonic oscillator case.  In what
follows we will report an experimental design in which the pacing
signal was completely absent. Moreover, we will consider conservative
parts of the CD modelling approach that correspond to
nonharmonic oscillators.

In the experiment
participants were sitting in a chair with their right arm fixed to a custom armrest,
allowing them to swing a pendulum (40~cm rod, cylindrical weight positioned
2.5~cm from end, 303~g total weight) with resonant frequency of 1~Hz
(6.28~rad/s) with their right hand about the wrist. Goniometers were
attached to the arm to capture the wrist angle (measured in radiants) while swinging the
pendulum (Biometrics, 100~Hz sampling rate).
For the duration of each one-minute trial, participants were instructed
to swing at one of three paces: comfortable, fast, or slow
(3 conditions, 3 repetitions each, presentation randomized). Rather
than synchronize their swinging to an external signal (i.e., a metronome)
for reference, participants determined their own pacing for all trials,
scaled relative to their natural (``comfortable'') swinging frequency.
Practice trials ensured that participants could generate 3 different
swinging paces and sustain them for one minute each. Practice began with
the ``comfortable'' condition. A tablet (Android, Nexus 10) was used
to approximate the participant pacing in beats per minute (BPMs), where
a ``beat'' was marked at the furthest point of the pendulum's arc.
Average BPM was also tracked for fast and slow conditions. ``Fast''
swinging was required to be at least 20\% faster than their
comfortable pace, and ``slow'' was likewise required to be at least
20\% slower in speed. If pacing during a slow or fast trial failed to
exceed this 20\% difference, participants were notified and prompted
to increase or decrease their pace accordingly. After practice,
most participants completed the experiment successfully and without
further corrections from the experimenter.
The experimental design allowed us to investigate and manipulate voluntary,
oscillatory human motor performance without the use of any external
pacing signal. To this end, the oscillatory goniometer signal was evaluated.

Let $x$ denote goniometer (wrist angle) signal in
radiants and let $v$ denote the corresponding velocity
$v=\rd x/\rd t$. Let
${\bf r}=(x,v)$ denote the state vector of the oscillatory system and
$H(x,v)$ a Hamiltonian function that will be specified later.
The Langevin equation~\cite{risken89book}
of the CD model for the
self-oscillator
reads~\cite{ebeling04book,frank05book,romanczuk12epjst}
\begin{equation}
\label{eq1}
\Dt {\bf r} = {\bf I} - \gamma \pmb\nabla_{v} \Phi +
\left(
\begin{array}{c}
0 \\
\sqrt{D} \Gamma(t)
\end{array}
\right),
\end{equation}
\looseness=-1 where
${\bf I} =
(\partial H/\partial v, \, -\partial H/\partial x)$ (from a dynamical system's
perspective) is the conservative
``force'' acting on the oscillator.
In equation~(\ref{eq1})
$\pmb\nabla_v=(0,\partial/\partial v)$ is the incomplete Nabla operator and
$\Phi=\Phi(H)$ is a function of $H$.
The parameter $D\geqslant 0$ denotes the diffusion coefficient and is composed of
the damping parameter $\gamma\geqslant 0$ and the noise amplitude $\theta\geqslant 0$ like
$D=\gamma \theta$. Finally, $\Gamma(t)$ denotes a Langevin
force~\cite{risken89book}
 normalized like
$\mean{\Gamma(t)\Gamma(t')}=2\delta(t-t')$. Here and in what follows
$\delta(z)$ represents the Dirac delta function
and
$\mean{\cdot}$ stands for ensemble averaging. In the conservative case (i.e.,
for $\gamma=0$ $\Rightarrow$ $D=0$) equation~(\ref{eq1}) corresponds to a Hamiltonian
dynamics. Therefore, we interpret $H$ as Hamiltonian function of the
oscillator.
The probability density of the state variables $x$ and $v$ is defined by
$P(x,v,t)=\left\langle\delta\big(x-x(t)\big)\delta\big(v-v(t)\big)\right\rangle$. From equation~(\ref{eq1}) it follows
that $P(x,v,t)$
satisfies the so-called free energy Fokker-Planck
equation~\cite{frank05book}
of the form
\begin{equation}
\label{eq2}
\dt P = - \pmb\nabla \cdot {\bf I} P + \pmb\nabla \cdot M \cdot P
\pmb\nabla \frac{\delta F_{\rm NE}}{\delta P} \, .
\end{equation}
In equation~(\ref{eq2}) we have the complete Nabla operator
$\pmb\nabla=(\partial/\partial x,\partial/\partial v)$ and
$\delta F_{\rm NE}/\delta P$ denotes the variational derivative of the
functional $F_{\rm NE}$ with respect to $P$.
The functional $F_{\rm NE}$ itself is one of the three
(non-equilibrium) thermodynamic variables
of the CD model and denotes the non-equilibrium free energy.
The two other variables are the non-equilibrium internal energy $U_{\rm NL}$
and the statistical entropy $S$ defined by~\cite{frank05book}
\begin{equation}
\label{eq3}
S= - \int P \ln P \, \mbox{d}x\mbox{d}v, \qquad
U_{\rm NL}=\mean{\Phi}, \qquad
F_{\rm NE}=
U_{\rm NL}
-  \theta S.
\end{equation}
Moreover, $M$ denotes a $2\times2$ mobility matrix with coefficients
$M_{22}=\gamma$ and $M_{jk}=0$ otherwise.
From equation~(\ref{eq3}) it follows that $\theta$
plays the role of a non-equilibrium
temperature. The
stationary solution of equation~(\ref{eq2})
in phase space $(x,v)$ reads
\begin{equation}
P(x,v) = \frac{1}{Z_{xv}}
\exp\left[
-\frac{\Phi\big(H(x,v)\big)}{\theta}
\right],
\end{equation}
where $Z_{xv}$ is a normalization constant.
Let us introduce the probability density of the Hamiltonian function
$H$ like
$P(H)=\int \delta\big(H-H(x,v)\big) \, P(x,v) \, \mbox{d}x\mbox{d}v$.
We obtain
\begin{equation}
P(H) = \frac{1}{Z_H}
\exp\left[
-\frac{\Phi(H)}{\theta}
\right],
\end{equation}
where $Z_H$ is a normalization constant again.
Note that the normalization constants $Z_{xv}$ and $Z_H$
are related to each
other~\cite{mongkolsakulvong10cmp,dotov15biosystems}.
Importantly,
$P(H)$ is maximal at the energy value $H$
for which $\Phi$ is minimal.
Therefore, $\Phi(H)$ can be regarded as
 non-equilibrium potential of $H$.
In addition, if we put $\Phi=(H-B)^2/2$, where $B$ is a parameter that will be
discussed below, then with respect to $P(H)$
the variables $F_{\rm NL}\,$,
$U_{\rm NL}$ and $S$ defined by equation~(\ref{eq3}) can be calculated from
$B$ and $\theta$ like~\cite{dotov15biosystems,kimseokhun15}
\begin{eqnarray}
\label{zus1}
U_{\text{NE}} =\frac{\theta}{2}
\left\{1 - \frac{B \exp[-B^2/(2\theta)]}{Z_H}\right\}, \qquad
S= \ln Z_H + \frac{U_H^{\text{NE}}}{\theta}\,, \qquad
F_{\text{NE}} = - \theta \ln Z_H
\end{eqnarray}
with
$Z_H=\sqrt{2\pi\theta} w(B/\sqrt{\theta})$, where $w(s)$
is the error function $w(s)=\int_{-\infty}^s \exp(-z^2/2) \mbox{d}z$.

Let us dwell on the interpretation of $\Phi$. Let us put $\gamma>0$ but
$\theta=0$ $\Rightarrow$ $D=0$ such that
\begin{equation}
\label{eq7}
\Dt x= \frac{\partial H}{\partial v}\,, \qquad
\Dt v = -\frac{\partial H}{\partial x}
- \gamma \frac{\partial \Phi}{\partial v} \qquad
\Rightarrow \qquad
\Dt \Phi = -
\gamma \left(\frac{\partial \Phi}{\partial v}\right)^2 \leqslant 0.
\end{equation}
That is, $\Phi$ is a nonincreasing function of time.
Frequently, the special case of a quadratic function
\begin{equation}
\Phi(H) = \frac{(H-B)^2}{2}
\end{equation}
has been discussed in the literature (e.g.~\cite{ebeling04book}).
In this case, equation~(\ref{eq7}) reads
\begin{equation}
\label{eq9}
\Dt x= \frac{\partial H}{\partial v}\,, \qquad
\Dt v = -\frac{\partial H}{\partial x}
- \gamma \frac{\partial H}{\partial v}(H-B)
\qquad \Rightarrow \qquad
\Dt H =
- \gamma \left(\frac{\partial H}{\partial v}\right)^2(H-B).
\end{equation}
\looseness=-1 Let us assume $H$ is bounded from below.
For the sake of simplicity we assume that (i)
$H\geqslant 0$ holds for all $x,v$, (ii)
$H$ is a smooth function defined on the phase space $x,v$,
and
there
exists at least one pair $x(\min),v(\min)$ such that $H=0$.
Furthermore, let us
assume that if $H>0$ holds then
trajectories $x(t),v(t)$ of equation~(\ref{eq9})
are such that $\partial H/\partial v=0$ holds only for
a discrete set $K$ of time points $t_1 <t_2 < t_3 \dots\;$.
Consequently, provided that $H>0$ and $H\neq B$ holds, then
we have $\mbox{d}H/\mbox{d}t<0$ if $t$ is not in the set $K$.
This implies that $H$ and $\Phi$ exhibit asymptotically
stable fixed points
at
$H=0$ and $\Phi=B^2/2$ for $B\leqslant 0$ and
 $H=B$ and $\Phi=0$ for $B>0$.
In particular, for both cases,
the potential $\Phi$ converges to its minimum value
$\Phi_{\min}=\min_H (\Phi)$.

So far, in
applications of the model~(\ref{eq1}) to
experimental research on human motor behavior,
the focus has been on the harmonic oscillator
case~\cite{dotov11mc,dotov15biosystems,kimseokhun15}
\begin{equation}
H= \frac{v^2}{2} + \frac{\omega^2 x^2}{2}\,.
\end{equation}
Taking
$B$ as a control parameter of the deterministic
oscillatory system~(\ref{eq1}) with $\theta=0$,
then there is a
Hopf bifurcation point at $B=0$. In the experiment, the parameter $B$
reflects the intention of a participant to swing ($B>0$) or not to
swing ($B\leqslant 0$) the pendulum. As far as the CD oscillator model is concerned,
for
$B\leqslant 0$ equation~(\ref{eq9}) exhibits an
asymptotically stable fixed point at $(x,v)=(0,0)$ with $H=0$.
By contrast, for
$B>0$ the fixed point $(x,v)=(0,0)$ is an unstable focus. There exists
a
stable limit cycle
characterized by the harmonic oscillator equation
$ {\rm d}^2 x /{\rm d} t^2  = -\omega^2 x$
and an oscillation amplitude $A$ given by
$\omega^2 A^2 /2 = B$~\cite{dotov11mc}.
Consequently, on the limit cycle we have
$x(t)=\omega^{-1}\sqrt{2B} \cos(\omega t + \varphi)$, where $\varphi$ is an
arbitrary phase.

Let us generalize the considerations in order to consider nonharmonic limit
cycles. To this end, we consider Hamiltonian functions of the form
\begin{equation}
\label{eq14}
H= \frac{v^2}{2} + V(x), \qquad
V(x)= V_0(x) + V_1(x).
\end{equation}
$V_0$ is considered as baseline oscillator potential and $V_1$ is considered as
a correction term. However, we do not require that $V_1$ should denote a small
correction term. By contrast, we require that the baseline potential should be
quadratic like
\begin{equation}
\label{eq15}
V_0(x) = k_1 x + k_2 \frac{x^2}{2}
\end{equation}
with $k_2>0$,
whereas $V_1$ accounts for
contributions different from the linear and quadratic cases.
In order to make sure that $V(x)$ is globally stable such that
non-equilibrium thermodynamic variables $F_{\rm NE}$ and $U_{\rm NL}$
and the statistical entropy
$S$ exist we require that $V_1(|x|\to \infty)=0$ should hold.
In what follows, we will use
\begin{equation}
\label{eq16}
V_1(x) = \exp(-x^2/2) \sum_{j=2}^N k_{j+1} x^{j+1}.
\end{equation}
\looseness=-1 In fact, for our purposes any other function $V_1$ could be used provided
 that $V_1(|x|\to \infty)=0$ holds and
$V_1$ is linearly independent with respect to
the linear and quadratic functions $x$ and $x^2$, respectively.
Let us consider the conservative case $\gamma=0$ $\Rightarrow$ $D=0$ again.
From $\mbox{d}{\bf r}/\mbox{d}t ={\bf I}$
and equations~(\ref{eq14}--\ref{eq16}) it then follows that
\begin{equation}
\frac{{\rm d}^2}{{\rm d} t^2} x = - k_1 - k_2 x -
\exp(-x^2/2) \sum_{j=2}^N k_{n+1} \left[
(n+1)x^n - x^{n+2}\right].
\end{equation}
Therefore, in the dissipative, deterministic case $\gamma>0$ with $\theta=0$
for appropriately chosen parameters $k_j$ with $N>2$,
the model~(\ref{eq9}) can exhibit stable limit cycles
that describe nonharmonic oscillatory behavior.
Importantly, in the fully dissipative case $\gamma>0$ with $\theta>0$,
the CD oscillator model (\ref{eq1}) can be used to determine
the non-equilibrium thermodynamic variables $F_{\rm NE}$ and $U_{\rm NL}$
and the statistical entropy
$S$ of the self-oscillator.

\looseness=-1 The variables $F_{\rm NE}\,$, $U_{\rm NL}\,$, and $S$ were estimated from
experimental data in a three-step approach. First, the model parameters
of the Hamiltonian $k_1,\ldots,k_n$ were determined using linear regression
analysis~\cite{silvap07,frank11ijmpb}. Second,
the canonical-dissipative parameters $\theta$ and
$B$ were estimated. To this end,
we followed
~\cite{dotov11mc,dotov15biosystems,kimseokhun15}
and calculated the Hamiltonian energy values $H$
from equations~(\ref{eq14}--\ref{eq16}) and
the experimentally observed time series $x(t)$ and $v(t)=\mbox{d}x/\mbox{d}t$
using the parameters estimates $k_1,\ldots,k_n\,$. Third,
following again the earlier work~\cite{dotov15biosystems,kimseokhun15},
we calculated
$F_{\rm NE}\,$, $U_{\rm NL}\,$, and $S$ from equation~(\ref{zus1}) and the estimated
values of
$\theta$ and~$B$.

\looseness=-1 A total of ten participants
(undergraduate students from the University of Connecticut who received
partial course credit for their participation) were tested.
Experimental procedures (as described above) were approved by the
University's Institutional Review Board (IRB).
Two participants performed
oscillatory movements that were subjected to a great amount of jerk and
could not be classified with the aforementioned data analysis procedure as
limit cycle oscillations. The data of those two participants were
discarded.
Only the data from the remaining
eight participants were evaluated. Two models were tested: the baseline model
involving only $V_0$ and the parameters $k_1$, $k_2$
(for the harmonic oscillator case) and the $N=5$ model
(for the nonharmonic oscillator case)
involving parameters $k_1,\ldots,k_6$ and terms up
to the order $x^6$.

\begin{table}[!b]
\caption{Angular frequencies $\omega$ and oscillation amplitudes $A$ as
functions of the three frequency (speed) conditions (slow, comfortable, fast).
Angular frequencies were determined as peak frequencies observed in the
Fourier spectrum of the movement trajectories $x(t)$. Amplitudes were
calculated from the difference
between averaged maximal and averaged minimal values of
$x(t)$: $A=[{\overline{ \rm Peaks(MAX)} }- {\overline{ \rm Peaks(MIN)} }]/2$.
Standard errors in parenthesis.}
\label{tab1}
\vspace{2ex}
%\begin{center}
\centering
\renewcommand{\arraystretch}{0}
\begin{tabular}{|l|c|c|c|}
\hline\hline
& Slow & Comfortable & Fast  \strut\\
\hline
\rule{0pt}{2pt}&&&\\
\hline
$\omega$ [rad/s] & 5.35 (0.36) & 7.01 (0.47) & 9.10 (0.71) \strut\\
\hline
$A$ [a.u.] & 0.27 (0.03) & 0.25 (0.03) & 0.28 (0.03) \strut\\
\hline\hline
\end{tabular}
\renewcommand{\arraystretch}{1}
%\end{center}
\caption{Improvements of model fits quantified by
$R^2$-improvements scores $\Delta R^2$ observed for the three
experimental frequency (speed) conditions.}
\label{tab2}
\vspace{2ex}
%\begin{center}
\centering
\renewcommand{\arraystretch}{0}
\begin{tabular}{|l|c|c|c|}
\hline\hline
& Slow & Comfortable & Fast \strut\\
\hline
\rule{0pt}{2pt}&&&\\
\hline
$\Delta R^2$ \permil & 1.6 & 2.6 & 1.4 \strut\\
\hline\hline
\end{tabular}
\renewcommand{\arraystretch}{1}
%\end{center}
\end{table}

Table~\ref{tab1} shows descriptive measures
(oscillation frequency and amplitude)
of the observed
oscillatory movements for the three frequency conditions.
As expected, oscillation frequency $\omega$
increased significantly
across the frequency conditions [$p<$0.01, F(2.14)=43.683].
The amplitude $A$ showed a U-shaped pattern with the
minimum at the comfortable condition. The effect of the frequency on $A$
was statistically significant [$p<$0.05, F(2.14)=5.045].
Table~\ref{tab2} shows the improvement $\Delta R^2$ of the $R^2$
goodness of fit scores when comparing
the nonharmonic case with the
harmonic case. By definition of the fitting
procedure, the $R^2$ scores of the more comprehensive model (nonharmonic case)
should
exceed the $R^2$ scores of the less comprehensive model
(harmonic case) such that $\Delta R^2>0$.
The $\Delta R^2$ scores reported in table~\ref{tab2} reflect this
fundamental feature of regression models.
For our purposes, table~\ref{tab2} demonstrates that we were able to improve
the fits of the self-oscillator models by taking nonharmonic contributions
into account.

Table~\ref{tab3} presents the bifurcation parameter $B$ as a function of
frequency
and model type. Consistent with the previous
studies~\cite{dotov11mc,dotov15biosystems,kimseokhun15},
there was a statistically significant
increase of $B$
from slow to fast oscillations speeds
[harmonic case: $p<$0.01,
F(2.14)=11.722; nonharmonic case: $p<$0.01, F(2.14)=10.694].
That is,
$B$ increased with increasing oscillation frequency.
Importantly, the
same pattern was qualitatively observed for both models.
Table~\ref{tab4} displays the statistical entropy $S$ and the
non-equilibrium thermodynamical variables $U_{\rm NL}$ and
$F_{\rm NL}$ for the models and frequency conditions. Entropy $S$ and internal
energy $U_{\rm NL}$ scores increased significantly as functions of
oscillation frequency
[harmonic case: $p<$0.01, F(2.14)=35.804 for $S$
and $p<$0.05, F(2.14)=3.991 for $U_{\rm NL}\,$;
nonharmonic case: $p<$0.01, F(2.14)=36.582 for $S$
and $p=$0.06, F(2.14)=3.503 for $U_{\rm NL}$]. Note that the increase of
$U_{\rm NL}$ in the nonharmonic case was only marginally statistically
significant.
 Consistent with the previous
studies~\cite{frank06pre,dotov15biosystems,kimseokhun15},
the non-equilibrium
free energy $F_{\rm NL}$ decreased when oscillation frequency increased. The
effect was marginally statistically significant
[harmonic case: $p=$0.07,
F(2.14)=3.188; nonharmonic case: $p=$0.09, F(2.14)=2.810].
We will return to this issue below.

%\textfloatsep=0mm

\begin{table}[!h]
\caption{Bifurcation parameter $B$ obtained for different frequency (speed)
conditions as obtained from the harmonic and nonharmonic model-based data
analysis. Standard errors in parenthesis.}
\label{tab3}
\vspace{2ex}
%\begin{center}
\centering
\renewcommand{\arraystretch}{0}
\begin{tabular}{|l||c|c|c|}
%\hline\hline
%Case & $B$(Slow) & $B$(Comfortable) & $B$(Fast) \strut\\
%& [a.u.] & [a.u.] & [a.u.] \strut\\
%\hline
%\rule{0pt}{2pt}&&&\\
%\hline
%Harmonic & 1.3 & 2.1 & 4.1 \strut\\
% &     (0.3) & (0.6) & (1.1)\strut\\
%\hline
%Non- & 1.3 & 2.2 & 4.5 \strut\\
%harmonic &      (0.3) &  (0.7) & (1.3) \strut\\
%\hline\hline
\hline\hline
Case & $B$(Slow) & $B$(Comfortable) & $B$(Fast) \strut\\
& [a.u.] & [a.u.] & [a.u.] \strut\\
\hline
\rule{0pt}{2pt}&&&\\
\hline
Harmonic & 1.3 (0.3)& 2.1 (0.6)& 4.1 (1.1)\strut\\
% &     & &\strut\\
\hline
Nonharmonic & 1.3 (0.3)& 2.2 (0.7)& 4.5 (1.3) \strut\\
%&      &  & \strut\\
\hline\hline
\end{tabular}
\renewcommand{\arraystretch}{1}
%\end{center}
\caption{Variables $S$, $U_{\rm NL}\,$, and $F_{\rm NL}$
as functions of the frequency (speed)
conditions (S=slow, C=com\-for\-table, F=fast) and
model type. Standard errors in parenthesis.}
\label{tab4}
\vspace{2ex}
%\begin{center}
\centering
\renewcommand{\arraystretch}{0}
\begin{tabular}{|l||c|c|c|c|c|c|c|c|c|c|c|c|}
\hline\hline
Model- & $S$(S) & $S$(C) & $S$(F) &
$U_{\rm NL}$(S) & $U_{\rm NL}$(C) & $U_{\rm NL}$(F) & $F_{\rm NL}$(S)
& $F_{\rm NL}$(C) & $F_{\rm NL}$(F) \strut\\
type & [a.u.] & [a.u.] & [a.u.] & [a.u.] & [a.u.]
& [a.u.] & [a.u.] & [a.u.] & [a.u.] \strut\\
\hline
\rule{0pt}{2pt}&&&&&&&&&\\
\hline
Harmonic & 1.1 & 1.3 & 1.9 & 0.5 & 0.9 & 2.9 &
-1.1 & -3.2 & -14.6\strut\\
 &  (0.2) & (0.2) & (0.3) & (0.1) & (0.5) & (1.4) &
(0.6) & (2.1) & (8.2)     \strut\\
\hline
Non- & 1.1 & 1.3 & 1.8 & 0.4 & 0.8 & 2.8 &
-1.0 & -2.8 & -13.9 \strut\\
harmonic &  (0.2) & (0.2) & (0.3) & (0.1) & (0.4) & (1.4) &
(0.5) & (1.9) & (8.1)    \strut\\
\hline\hline
\end{tabular}
\renewcommand{\arraystretch}{1}
%\end{center}
\end{table}

Our experimental design allowed us to examine and manipulate voluntary
 human performance in an oscillatory motor control task without the use of
an external pacing signal. Moreover, we were able to account for
nonharmonic oscillator contributions in the conservative part of the
CD modelling approach. As expected, the nonharmonic
models matched the data better than the harmonic models. Importantly, the
thermodynamic state variables showed for the harmonic and nonharmonic
 analyses the same kind of patterns across the
three frequency conditions. Therefore,
 it seems that (at least for the experimental paradigm of the present
study) the thermodynamic variables are not affected by the model type.

The pumping parameter $B$ increased with oscillation frequency. It has been
suggested that $B$ measures the distance to the Hopf bifurcation
point of the canonical-dissipative
self-oscillator~\cite{dotov15biosystems,kimseokhun15}.
Accordingly, the
non-equilibrium free energy decayed when the tested ``human'' self-oscillators
operated further away from their bifurcation points.
It has been argued that this feature is consistent with the decay of the
Landau free energy of magnetic phase transitions~\cite{strobl04book}
when the temperature is
decreased further below the critical transition
temperature~\cite{dotov15biosystems,kimseokhun15}.

For the sample of eight participants, the decay of the non-equilibrium
free energy was only marginally statistically significant. However, a detailed
inspection of the individual participant data revealed that two of the
eight participants showed free energy scores $F_{\rm NL}$
that were by a factor of
10 larger than the scores obtained for the remaining six participants.
Analyzing the non-equilibrium free energy only for the remaining
``homogeneous'' group of six participants, we found for $F_{\rm NL}$ the same
qualitative pattern as for the whole group: the free energy decayed
 with an increasing oscillation frequency. However, for this homogeneous
subset of participants, the effect of speed was also statistically significant
($p<$0.05 for both models). Future experimental work may
be conducted to explore this participant effect in more detail.

\vspace*{-2ex}%
\section*{Acknowledgements}
\vspace*{-1ex}%
Preparation of this manuscript was supported in part by
National Science Foundation under the INSPIRE track,
grant BCS-SBE-1344275.

%\vspace{-1mm}
\ukrainianpart
\title{Лінійна нерівноважна термодинаміка людської поведінки:
метод канонічно-дисипативного рівняння Фоккера-Планка з потенціалами, що включають  ангармонізми
}
\author[Дж.М. Гордон, С.Кім, Т.Д. Франк]{Дж.М. Гордон\refaddr{label1},
С.Кім\refaddr{label1},
Т.Д. Франк\refaddr{label1,label2}}
\addresses{
\addr{label1}
CESPA, факультет психології, Університет Коннектикуту, Сторрс,
CT 06269, США\\
\addr{label2}
Факультет фізики, Університет Коннектикуту, Сторрс,
CT 06269, США
}
\vspace{-5mm}
\makeukrtitle

\begin{abstract}
Представлено нову експериментальну парадигму та новий підхід до моделювання роботи осциляторного  людського двигуна
з допомогою ключового концепту фізики конденсованої речовини, а саме --- змінних термодинамічного стану.
Для цього в новій експериментальній парадигмі учасники здійснювали маятникові коливні рухи з особисто вибраними частотами, що
відмінне  від попередніх досліджень, де використовувалися синхронізуючі сигнали. У новому підході моделювання
використано канонічно-дисипативну границю циклічно осциляційної моделі з консервативною частиною, що враховує негармонічні
коливні компоненти --- у цьому інша відмінність від попередніх досліджень, у яких розглядалися лише гармонічні компоненти.
В узгодженні з теорією Ландау магнітних фазових переходів ми знайшли, що вільна енергія осциляційної моделі зменшується, коли учасники
підтримують коливання з віддаленням від точки біфуркації Хопфа для канонічно-дисипативної границі циклічного осцилятора.

\keywords фізика життя, людська поведінка, канонічно-дисипативні системи, змінні термодинамічного стану
\end{abstract}
\end{document}